\pdfoutput=1
\documentclass[conference]{IEEEtran}
\usepackage[T1]{fontenc}
\usepackage{amsmath}
\usepackage{amsfonts}
\usepackage{amssymb}
\usepackage{cite}
\usepackage{multicol}
\usepackage{verbatim}
\usepackage{graphicx}
\usepackage{subcaption}
\usepackage{balance}
\usepackage{epstopdf}
\usepackage{epsfig,color}
\usepackage{subcaption}
\usepackage[linesnumbered,ruled]{algorithm2e}

\SetKwInOut{Parameter}{Parameters}
\SetKwRepeat{Do}{do}{while}

\IEEEoverridecommandlockouts

\begin{document}



\title{Maximizing Secrecy Rate of an OFDM-based Multi-hop Underwater Acoustic Sensor Network}

\author{
\IEEEauthorblockN{ Waqas Aman$^\ast$, 
M. Mahboob Ur Rahman$^\ast$
, Zeeshan Haider$^\ast$
, Junaid Qadir$^\ast$, M. Wasim Nawaz$^\bot$, \\ and Guftaar Ahmad Sardar Sidhu$^\dagger$ 
} 
\IEEEauthorblockA{
$\ast$Department of Electrical engineering, Information Technology University, Lahore, Pakistan\\ 
$\bot$Department of Computer engineering, The University of Lahore, Lahore, Pakistan\\ 
$\dagger$Department of Electrical engineering, COMSATS University Islamabad, 45500 Islamabad, Pakistan \\
$^\ast$\{waqas.aman,mahboob.rahman,junaid.qadir\}@itu.edu.pk, $^\bot$muhammad.wasim@dce.uol.edu.pk, $^\dagger$guftaarahmad@comsats.edu.pk }
}

\maketitle 


\begin{abstract}

In this paper, we consider an eavesdropping attack on a multi-hop, UnderWater Acoustic Sensor Network (UWASN) that consists of $M+1$ underwater sensors which report their sensed data via Orthogonal Frequency Division Multiplexing (OFDM) scheme to a sink node on the water surface. Furthermore, due to the presence of a passive malicious node in nearby vicinity, the multi-hop UnderWater Acoustic (UWA) channel between a sensor node and the sink node is prone to eavesdropping attack on each hop. Therefore, the problem at hand is to do (helper/relay) node selection (for data forwarding onto the next hop) as well as power allocation (across the OFDM sub-carriers) in a way that the secrecy rate is maximized at each hop. To this end, this problem of Node Selection and Power Allocation (NSPA) is formulated as a mixed binary-integer optimization program, which is then optimally solved via decomposition approach, and by exploiting duality theory along with the Karush-Kuhn-Tucker conditions. We also provide a computationally-efficient, sub-optimal solution to the NSPA problem, where we reformulate it as a mixed-integer linear program and solve it via decomposition and geometric approach.  Moreover, when the UWA channel is multipath (and not just line-of-sight), we investigate an additional, machine learning-based approach to solve the NSPA problem. Finally, we compute the computational complexity of all the three proposed schemes (optimal, sub-optimal, and learning-based), and do extensive simulations to compare their performance against each other and against the baseline schemes (which allocate equal power to all the sub-carriers and do depth-based node selection). In a nutshell, this work proposes various (optimal and sub-optimal) methods for providing information-theoretic security at the physical layer of the protocol stack through resource allocation.

\end{abstract}

\section{Introduction}

UnderWater Acoustic Sensor Networks (UWASNs) find their utilization by a multitude of civilian, commercial and military applications, e.g., marine life exploration, intrusion detection for border surveillance, performance monitoring of oil rigs, searching for (oil, gas, minerals) resources underwater, to name a few \cite{Ian:AdhocNet:2005},\cite{Felemban:IJDSN:2015}. Contrary to the terrestrial communication, UnderWater Acoustic (UWA) communication is quite challenging because the UWA channel is characterized by frequency-dependent pathloss, colored Gaussian noise, low symbol-rate due to long propagation delays (due to low speed of acoustic waves underwater), and fading effects due to multipath propagation \cite{Ian:AdhocNet:2005},\cite{Chen:Network:2018}.

In addition to the aforementioned challenges, UWA channel---being a broadcast channel---is also susceptible to various kinds of attacks by the active and passive adversaries nearby \cite{Jiang:COMMST:2018},\cite{Han:CommMag:2015}. To this end, like their terrestrial counterparts, UWASNs have traditionally been secured by employing cryptographic measures at the higher layers of the protocol stack. But unfortunately, such measures are  fallible: a concrete example is the work \cite{Zhang:IWC:2002} where authors demonstrated that they could break into the crypto-based security measures employed by the IEEE 802.11/Wi-Fi systems. 
Thus, the vulnerability of the cryptography-based security measures against the brute-force attacks that could be launched by the adversaries has prompted the researchers to find alternate as well as complementary mechanisms of securing the UWASNs. To this end, a set of techniques under the umbrella term Physical Layer Security (PLS) has received considerable attention by the researchers lately \cite{Bloch:Book:2011},\cite{Zhou:Book:2013}. PLS exploits the random nature of physical propagation medium to provide information security, and thus operates at the physical layer of the protocol stack.

The existing literature on the PLS techniques could be broadly classified into two main categories: i) works which provide information-theoretic bounds on the performance of communication systems under attack \cite{Bloch:Book:2011},\cite{Zhou:Book:2013},\cite{Wyner:BSTJ:1975}, and ii) the works which present algorithms that exploit the features at the physical layer (medium-based \cite{Ammar:VTC:2017}, or, hardware-based \cite{Waqas:VTC:2019},\cite{Mahboob:VTC:2017},\cite{Mahboob:Globecom:2014}) as device fingerprints to ensure security and trust among the legitimate nodes. Having said that, this work belongs to the first category of the works, i.e., the information-theoretic PLS. 

A large body of the works on Information-theoretic PLS computes the so-called secrecy rate\footnote{Secrecy rate (secret bits/sec) is defined as the rate between two legitimate nodes minus the leakage to eavesdropper \cite{Wyner:BSTJ:1975}.} for various system models, configurations of interest and discusses ways to maximize it. For example, \cite{Waqas:WCNC:2016} and \cite{Waqas:ett:2018} maximize secrecy rate through joint optimization of carrier and power allocation with relay selection in Orthogonal Frequency Division Multiplexing (OFDM) based cooperative communication networks. \cite{Izanlou:ett:2019} maximizes the secrecy rate in a device-to-device communication link through optimal power allocation. Last but not the least, \cite{Mashdour:ett:2020} utilizes artificial noise along with optimal power allocation to maximize the secrecy rate in a millimeter-wave communication link.

Inline with previous works on Information-theoretic PLS, this work maximizes the secrecy rate of a multi-hop UWASN. Specifically, this work presents novel (optimal and sub-optimal) methods to solve the problem of Node Selection and Power Allocation (NSPA) across the OFDM sub-carriers such that the secrecy rate at each hop is maximized. Additionally, this work also compares the computational complexity of all the proposed schemes. 
In simulations, we compare the performance of all the proposed (optimal and sub-optimal) schemes against a baseline scheme (i.e., depth-based node selection scheme). We notice that the secrecy rate provided by sub-optimal schemes is low, but their computational complexity is also lesser compared to the optimal schemes. Finally, we note that an increase in the power budget leads to an increase in the secrecy rate, and vice versa.

{\bf Outline.} The rest of this paper is organized as follows. Section II provides a compact summary of the selected related works, outlines the research gap, and lists the contributions of this work. Section III describes the system model as well as the two UWA channel models, i.e., Line-of-Sight (LoS) and multipath. Section IV (Section V) presents in detail the proposed optimal solution (sub-optimal solution) to the NSPA problem, for the LoS UWA channel. Section VI presents the proposed optimal, sub-optimal and machine learning-based solutions to the NSPA problem, for the multipath UWA channel. Section VII provides detailed simulation results. Section VIII concludes the paper.

\section{Related work \& Contributions of This Work} 

Even though the security challenges faced by the terrestrial networks are well-studied and corresponding crypto-based solutions are well-investigated, the literature addressing the security needs and solutions for UWASNs is relatively scarce (see the survey articles \cite{Han:CommMag:2015}, \cite{Domingo:WC:2011}, \cite{Cong:ICCMC:2010}, \cite{Lal:JOE:2017}, and the references therein). The works \cite{Han:CommMag:2015}, \cite{Domingo:WC:2011,Cong:ICCMC:2010,Lal:JOE:2017} unanimously state that numerous kinds of attacks could be launched onto the UWASNs, e.g., impersonation (or, intrusion) attacks, eavesdropping attacks, Sybil attacks, denial-of-service attacks, wormhole attacks, jamming attacks, man-in-the-middle attacks, and malicious relaying, to name a few. Accordingly, we cluster/group together the works that address similar kind of attacks below.

\subsection{Attacks on UWASNs \& Countermeasures}

{\it Crypto-based authentication.} 
The works \cite{DiniCC:ISCC:2011},\cite{DiniND:ISCC:2011} and \cite{Spaccini:OCEANS:2015} all propose various cryptographic measures to realize secure communication among the members of a UWASN. Specifically, \cite{DiniCC:ISCC:2011} suggests that a dedicated sink node distributes and manages pre-defined group keys and session keys to the UWASN members in order to counter eavesdropping attack and impersonation attack by the malicious nodes. In \cite{DiniND:ISCC:2011}, authors consider spoofing attacks and denial-of-service attacks on a UWASN during the network discovery phase, and propose modifications (i.e., formation of clusters and distribution of cluster keys to the members) to a well-known network discovery protocol (i.e., the so-called FLOOD protocol) to keep the network discovery phase secure. Finally, \cite{Spaccini:OCEANS:2015} proposes to utilize symmetric keys and  asymmetric keys for authentication and message encryption within a UWASN.

{\it Shared secret key generation.} 
Shared secret key generation is a classical problem within the domain of PLS whereby a legitimate node pair extracts secret keys from a mutual random source (typically, some characteristic of the underlying communication channel). For UWASNs, works \cite{Liu:ICSP:2008}, \cite{Huang:TWC:2016} generate shared secret keys by exploiting the unique characteristics of the UWA channel. To be more concrete, \cite{Liu:ICSP:2008} considers a reciprocal multipath time-varying acoustic channel and generates shared secret keys by exploiting the received signal strength as a mutual random source. On the other hand, \cite{Huang:TWC:2016} generates the shared secret keys from the frequency response of the acoustic channel. On a slightly different note, \cite{Xu:TMC:2018} presents SenseVault whereby the authors suggest that a UWASN may be divided into many clusters. Then, for each cluster, authors propose to generate and manage cryptographic/hash-based secret keys for authentication of messages that could either be received from the cluster members, or, from the nodes in other clusters. 

{\it Secure routing.} 
The recent surge of interest in multi-hop UWASNs has prompted the researchers to design a plethora of routing protocols, each with a different design objective, to address the unique set of challenges that UWASNs pose. (see the survey article \cite{Ning:Sensors:2016},\cite{Waqas:AINA:2016}, \cite{Bu:ett:2018} and the references therein for more details).
There are few publications on secure routing in UWASNs \cite{Zhang:INFOCOM:2010} and \cite{Xiujuan:IJDSN:2017}, which design routing protocols to detect a wormhole link in an UWASN. In \cite{Zhang:INFOCOM:2010}, Zhang et. al. detect a wormhole link through a set of neighbor discovery protocols based on the direction of arrival of the acoustic wave. A secure, anonymous routing protocol is presented in \cite{Xiujuan:IJDSN:2017} which performs two-way signature-based authentication under the assumption that the attacker node has no information (such as location, ID, etc.) about the legitimate nodes.

{\it Active attacks.} The works \cite{Goetz:IWUWN:2011,Zuba:SCN:2015,Xiao:Globecom:2015} study the jamming attacks, while the works \cite{Waqas:Access:2018,Waqas:ICC:2020} study impersonation attacks on UWASNs. Jamming attacks first. \cite{Goetz:IWUWN:2011} suggests the idea of restricted flooding whereby the data of a sensor node is sent to the sink node via multiple multihop paths with the hope that it makes the UWASN jamming-resilient. \cite{Zuba:SCN:2015} presents the findings of the real-time jamming experiments conducted by the authors in Mansfield Hollow Lake (in Mansfield, CT, USA)---that is, jamming attacks on a UWASN culminate in denial-of-service dilemma. \textit{Xiao et al.} In \cite{Xiao:Globecom:2015}, authors perform a game-theoretic analysis of the jamming attack in order to provide closed-form expressions for the Nash equilibrium for the case when all the UWA channels are known. For the case of unknown UWA channels, they utilize reinforcement learning to implement transmit power control. Next, the impersonation attacks. \cite{Waqas:Access:2018} considers impersonation attack on a UWASN by multiple malicious nodes and thwarts it by utilizing a two-step authentication procedure (whereby the sink node implements a distance-bounding test that is followed by another angle-of-arrival based hypothesis test). In \cite{Waqas:ICC:2020}, the authors compute the so-called effective capacity in order to quantify the reliability/quality-of-service performance of a UWA channel that is under threat of impersonation attack by a malicious node nearby.

{\it Passive attacks.} Eavesdropping attacks on UWASNs are studied in \cite{Dai:Sensors:2016} and \cite{Huang:SJ:2016}. Specifically, \cite{Dai:Sensors:2016} considers a single eavesdropper Eve that attempts to overhear the ongoing communication between the nodes of a UWASN, and computes the success probability of Eve (i.e., the probability that Eve can decode what she hears). In\cite{Huang:SJ:2016}, the authors maximize the secrecy rate of a UWA channel that is under threat by a single eavesdropper Eve. Specifically, they suggest that the receiver node transmits a well-crafted noise-like signal that combines with transmitter's signal at Eve (which equivocates Eve)---thanks to the large propagation delays of the UWA channel.


{\it Research Gap.} In contrast to the related work summarized above, this work considers the problem of data forwarding from a sensor node to the sink node when a passive malicious node (Eve) is present in the close vicinity. To the best of authors' knowledge, {\it this problem of maximizing the secrecy rate of an OFDM-based multi-hop UWASN has not been considered in the literature before}.

\subsection{Contributions of This Work}
The main contributions of this work are:

\begin{itemize}
\item The NSPA problem is formulated as a mixed binary-integer optimization program, which is then optimally solved via decomposition approach, and by exploiting duality theory along with the Karush-Kuhn-Tucker (KKT) conditions. We provide a computationally-efficient, sub-optimal solution to the NSPA problem, where we reformulate it as a mixed-integer linear program and solve it via decomposition and geometric approach.  
\item When UWA channel is multipath (and not just LoS), we investigate an additional Machine-Learning (ML) based approach to solve the NSPA problem. Finally, we compute the computational complexity of all the three proposed schemes (optimal, sub-optimal, and learning-based). Specifically, learning based architecture is proposed which comprises two Neural Networks (NNs).
\end{itemize}

\section{System Model \& UWA Channel Models}

\subsection{System Model}
\begin{figure}
\begin{center}
\includegraphics[width=9cm]{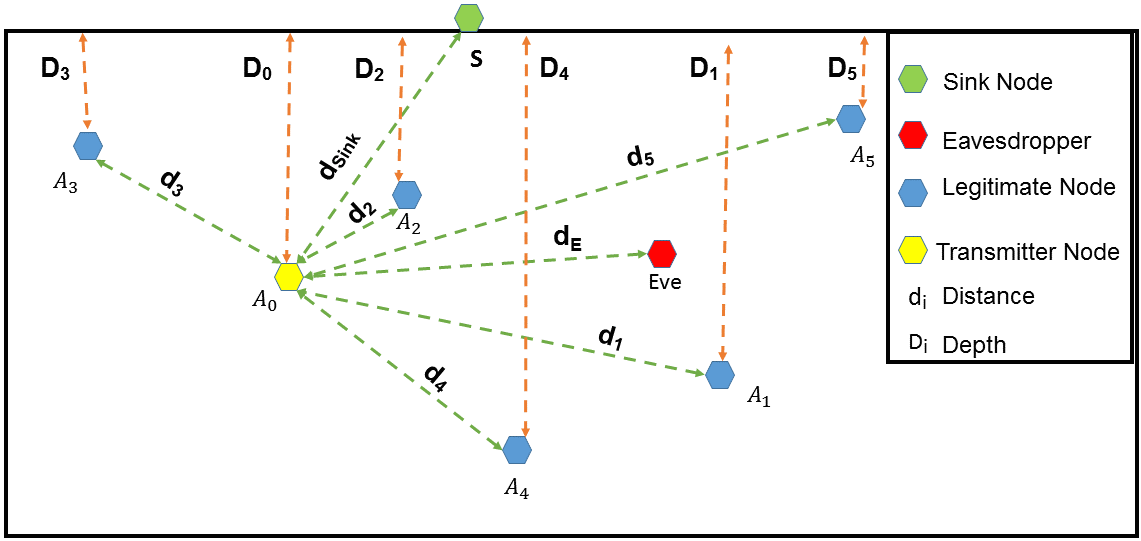}
\caption{The system model.}
\label{fig:system-model}
\end{center}
\end{figure}

We consider a UWASN comprising $M+1$ underwater sensor nodes (so-called Alice nodes) that report their sensed data (via OFDM scheme with $N$ sub-carriers) to a sink node $S$ on the water surface (see Fig. \ref{fig:system-model}). Let $\mathcal{A}=\{A_0,A_1,...,A_M\}$ represent the set of Alice nodes. Without loss of generality, we assume that the node $A_0$ has to report its sensed data to $S$. Let $D_i$ represent the depth of the node $A_i$ from the water surface, while $d_i$ represents the distance of node $A_i$ from $A_0$. We assume that a {\it passive} eavesdropper (Eve) is present in the close vicinity of the UWASN. We further assume that the Eve is not very closely located to the sink node\footnote{This assumption is needed for the proposed algorithms (to be described in the next section) to terminate. This assumption is reasonable because the sink node is typically a very powerful node equipped with proximity sensors (and thus, is capable of detecting a malicious node nearby).}. 

We consider two kinds of UWA channel models in this work.

\subsection{LoS UWA Channel Model}
The frequency-dependent path-loss between a transmit node and a receive node with separation $d$ is given (in dB scale) as \cite{Stojanovic:1ACMIWUN:2006}:
\begin{align}
\label{eq:pl}
PL(d,f)_{dB} = \nu 10\log d + d \alpha(f)_{dB}, 
\end{align}
where $\nu$ is the so-called spreading factor, while $\alpha(f)$ is the coefficient of absorption given as:
\begin{align}
\alpha(f)_{dB}=\frac{0.11            
f^2}{1+ f^2}+\frac{44f^2}{4100+f^2}+2.75\times 10^{-4}f^2+0.003.
\end{align}
$N(f)$ is the frequency-dependent Power Spectral Density (PSD) of the ambient noise (comprising of noise contributions from turbulence, shipping, waves, and thermal noise) \cite{Stojanovic:1ACMIWUN:2006}:  
\begin{align}
\label{eq:noise}
N(f)_{dB} \approx N_1- \tau 10\log f,  
\end{align} 
where $N_1$ and  $\tau$ are the experimental constants. Note that the above approximation of the PSD $N(f)$ of ambient noise holds for frequency range $1-100$ kHz only \cite{Stojanovic:1ACMIWUN:2006}. 

\subsection{Multipath UWA Channel Model}
We consider a UWA channel described by its transfer function $H_f$ where $f$ is the frequency. Let $H_i$ be the value of $H_f$ at the center frequency of $i^{th}$ sub-carrier. We assume that $H_i$ is flat in the band $\Delta f$. We consider a multipath UWA channel with total $L$ paths. Accordingly, the UWA channel gain during $k^{th}$ block/time-slot is as follows: 
\begin{equation}
\label{CGinMP}
H_i(k)=\sum_{l=1}^L \frac{1}{\sqrt{PL(d_l,f)}} h_l(k) e^{-j2\pi f_i \tau_l(k)}, 
\end{equation}
where $d_l$ is the distance covered and $\tau_l(k)=\frac{d_l}{v}$ is the time taken by $l^{th}$ path, where $v$ is the speed of acoustic wave under the water.
The path gains $h_l(k)$ are modeled as independent, first-order auto regressive process \cite{Aval:Ucomms:2014}.
\\

\section{Optimal Solution to the NSPA problem in LoS UWA channel}

We consider a situation where the direct link between $A_0$ and $S$ does not exist. Therefore, the problem at hand is to design a scheme to route the data of $A_0$ to $S$, while protecting the data (at each hop) as much as possible from the eavesdropping attack by Eve. To this end, we propose that $A_0$ should forward its data to that relay node $A_i\in \mathcal{C}$ ($\mathcal{C} = \mathcal{A}\cup S \setminus A_0$) whose Secrecy Rate (SR) is maximum among all other candidate helper nodes. The SR of $A_i$ (summed over all the $N$ OFDM sub-carriers) is defined as follows:
\begin{align}
\mathrm{SR}^{(i)}=\bigg( \Delta f (\sum_{j=1}^{N} \log_2(1+\mathrm{SNR}_{j}^{(i)})-\log_2(1+\mathrm{SNR}_{j}^{(E)})) \bigg)^{+} \label{obj},
\end{align}
where $(x)^+=\max(x,0)$; $\mathrm{SNR}_{j}^{(i)}$ ($\mathrm{SNR}_{j}^{(E)}$) is the Signal-to-Noise Ratio (SNR) on $j$-th sub-carrier at $A_i$ (Eve), and $\Delta f$ is the bandwidth of a sub-carrier (or, the spacing between the sub-carriers). 
The SNR on the $j$-th sub-carrier at $A_i$ is:
\begin{align}
\mathrm{SNR}_{j}^{(i)}=\frac{p_j}{(\int_{(j-1)\Delta f}^{j\Delta f}PL^{(i)}(f)df) (\int_{(j-1)\Delta f}^{j\Delta f}N(f) df)}.
\end{align}
Similarly, the SNR on the $j$-th sub-carrier at Eve is:
\begin{align}
\mathrm{SNR}_{j}^{(E)}=\frac{p_j}{(\int_{(j-1)\Delta f}^{j\Delta f}PL^{(E)}(f)df) (\int_{(j-1)\Delta f}^{j\Delta f}N(f) df)},  
\end{align}
where $p_j$ is the transmit power of $A_0$ over the $j$-th sub-carrier. 

To select one such node for data forwarding whose secrecy rate is maximum among all other candidate helper nodes, $A_0$ formulates the following optimization problem:
\begin{align}
&\max_{(\{\eta_i\}_{i\in \mathcal{C}}, \{p_j\}_{j=1}^N)} \sum_{i\in \mathcal{C}} \eta_i \mathrm{SR}^{(i)}(\{p_j\}_{j=1}^{N})   \label{eq:opt}\\
\text{s.t.} \ \  &\sum_{i\in \mathcal{C}}  \eta_i \sum_{j=1}^N p_j\leq P_T \nonumber\\
		&\sum_{i\in \mathcal{C}} \eta_i= 1   \nonumber  
\end{align}
where $\eta_i \in \{0,1\}$ $\forall i \in \mathcal{C}$; $\eta_i=1$ ($\eta_i=0$) implies that the helper node $A_i$ is selected (not selected). The first constraint of the optimization problem  (\ref{eq:opt}) ensures that for any candidate helper node $A_i$, the total power allocated over the $N$ sub-carriers should not exceed the total power budget $P_T$ of the sensor node $A_{0}$. The second constraint ensures that only one node is selected for data forwarding at each hop.

The optimization program  (\ref{eq:opt}) is a mixed binary-integer program; we adopt a decomposition approach to solve it.
For any selected node the problem decomposes to
\begin{align}
&\max_{\{p_j\}_{j=1}^N} \mathrm{SR}^{(i)}(\{p_j\}_{j=1}^{N})   \label{eq:subopt1}\\
\text{s.t.} \ \  &\sum_{j=1}^N p_j\leq P_T \nonumber \\
& p_j \geq 0 \ \ \forall \ \ j 	 \nonumber
\end{align}
The problem (\ref{eq:subopt1}) is now a convex optimization program and we exploit duality theory for getting a solution with zero duality gap. The associated dual program can be written as
\begin{align}
&\min_{\lambda} \ \ \max_{\{p_j\}_{j=1}^N}L(\{p_j\}_{j=1}^N)  \label{eq:Dual subopt1}\\
\text{s.t.} \ \  &\lambda \geq 0 \nonumber
\end{align}  
where $L(.)$ is the Lagrangian function and $\lambda$ is the dual variable.\\
The program (\ref{eq:Dual subopt1}) allows us to first solve the inner maximization. We use dual decomposition to get the solution for inner maximization. The decomposed problem can be written as
\begin{align}
&\max_{p_j\geq 0} \ \ \log_2\left(\frac{\Omega_j^{(i)} \Omega_j^{(E)}+p_j\Omega_j^{(E)}}{\Omega_j^{(i)} \Omega_j^{(E)}+p_j\Omega_j^{(i)}}\right)+\lambda P_T - \lambda p_i \label{scnv}  
\end{align}
where $\Omega_j^{(i)}=\int_{B_j(f)}PL^{(i)}(d,f)df . \int_{B_j(f)}N(f) df$, and $\Omega_j^{(E)}=\int_{B_j(f)}PL^{(E)}(d,f)df . \int_{B_j(f)}N(f) df$; $\Omega_j^{(i)}$ ($\Omega_j^{(E)}$) is the net noise power observed by $A_i$ (Eve). The expression in the argument of $\log_2(.)$ above is obtained through logarithmic property, i.e., $\log a-\log b=\log (\frac{a}{b})$. 
The Lagrangian associated with Eq. (\ref{scnv}) is:
\begin{align}
\Delta=\log_2\left(\frac{\Omega_j^{(i)} \Omega_j^{(E)}+p_j\Omega_j^{(E)}}{\Omega_j^{(i)} \Omega_j^{(E)}+p_j\Omega_j^{(i)}}\right)+\lambda (P_T-p_j) + \mu_jp_j,  \nonumber
\end{align}
where $\mu_j$ is the Lagrangian multiplier associated with $p_j$. Now, exploiting KKT  conditions, we obtain
 the following solution: 
\begin{align}
\label{eq:optpower}
p_j^*=\left(\frac{-b_j+\sqrt{b_j^2-4a_jc_j}}{2a_j}\right)^+,
\end{align}
where $a_j=\Omega_j^{(i)}\Omega_j^{(E)}$,\ \ $b_j=\Omega_j^{2(i)}\Omega_j^{(E)}+\Omega_j^{(i)}\Omega_j^{2(E)}$ \\ and  $c_j=(\Omega_j^{(i)}\Omega_j^{(E)})^2-\frac{\Omega_j^{(i)}\Omega_j^{2(E)}}{\lambda\ln(2)}+\frac{\Omega_j^{2(i)}\Omega_j^{(E)}}{\lambda\ln(2)}$.\\
After getting all $p_j*$ and putting back to program (\ref{eq:Dual subopt1}), we are left with external optimization problem given as:
\begin{align}
&\min_{\lambda}  L(\{p_j^*\}_{j=1}^{N}) \label{Dual:subproblem1}   \\
\text{s.t.} \ \  
&\lambda \geq 0 \nonumber
\end{align} 
To solve problem (\ref{Dual:subproblem1})  we use the sub-gradient method, which iteratively solves problem (\ref{Dual:subproblem1}) according to the following control law: 
\begin{align}
\label{eq:controllaw}
\lambda(m+1)= \lambda(m)+\delta(P_T-P_{\mathrm{alloc}}(m)),
\end{align}
where $\delta$ is the step size, and $P_{\mathrm{alloc}}(m)= \sum_{j=1}^N p_j^*(m)$. The algorithm converges when $P_{\mathrm{alloc}}(m)=P_T$. This completes solution to the problem (\ref{eq:subopt1}).\\
Now, we are left to solve the remaining problem which can be expressed as:

\begin{align}
&\max_{\{\eta_i\}_{i\in \mathcal{C}}} \sum_{i\in \mathcal{C}} \eta_i \mathrm{SR}^{(i)}(\{p_j^*\}_{j=1}^{N})   \label{eq:optb}\\
\text{s.t.} \ \  &\sum_{i\in \mathcal{C}} \eta_i= 1   \nonumber
\end{align}
The program (\ref{eq:optb}) is a binary program. Let $i^*=\arg \underset{i}\max$ $\mathrm{SR}^{(i)}(\{p_j^*\}_{j=1}^{N})  \ \  \forall i \in \mathcal{C}$, then the optimal solution to problem (\ref{eq:optb}) is:
\begin{align}
\label{eq:gammastar}
\eta_i^* &=\begin{cases} 
1, &i=i^* \ \ \& \ \ D_i < D_0 \\
0 &\ \mathrm{else}
\end{cases}
\end{align}
where $D_i$ is the depth of $A_i \ \ \forall i \in \mathcal{C}$ and $D_0$ is the depth of transmitter node.

The proposed method (when run on the first hop) is fully summarized in Algorithm 1. 
The Algorithm 1 is repeatedly invoked at each helper node to select the node for the next hop until the data reaches the sink node.

\begin{algorithm}
    \SetKwInOut{Input}{Input}
    \SetKwInOut{Output}{Output}

    \Input{${d_i, D_i \ \forall i, d_E}$}
    \Output{$p_j^* \ \forall j, \eta_i^* \  \forall i$}
    \Parameter{ $\lambda(0),\delta, P_T, M, N$ }
    \underline{Optimization:} \\
    
    \While{(1)}{
    \Repeat{$P_{\mathrm{alloc}}=P_T \  \forall i$}{
      implement Eq. (\ref{eq:optpower}) $\forall j$ \;
      implement Eq. (\ref{eq:controllaw}) \;
    }
    }
    return $p_j^*,\forall j$ \;  
    implement Eq. (\ref{eq:gammastar}) to return $\eta_i^*$ \;
    \caption{The proposed optimal scheme for NSPA problem (for LoS UWA channel)}
\end{algorithm}
\begin{algorithm}
    \SetKwInOut{Input}{Input}
    \SetKwInOut{Output}{Output}

    \Input{${d_i, D_i \ \forall i, d_E}$}
    \Output{$\mathbf{p^*}, \mathbf{\eta^*} $}
    \Parameter{ $ P_T, M, N$ }
    \underline{Optimization:} \\
%
      implement Eq. (\ref{eq:optLP}) to return $\mathbf{p^*}$ \;
      implement Eq. (\ref{eq:opt6}) to return $\mathbf{\eta^*}$ \;
      end
    
    \caption{The proposed sub-optimal scheme for NSPA problem (for LoS UWA channel)}
\end{algorithm}
\section{Sub-optimal Solution to the NSPA problem in LoS UWA channel}
The solution presented in the previous section is an iterative approach which takes a finite number of iterations before convergence. Therefore, in this section, we provide a one-shot, sub-optimal solution to the NSPA problem. Specifically, instead of maximizing the original objective function, we maximize the SNR difference of the $i$-th legal node and Eve. Thus, we formulate a new objective function for any $i$-th node as: $\sum_{j=1}^N\left(\frac{\Omega_j^{(E)}-\Omega_j^{(i)}}{\Omega_j^{(i)}\Omega_j^{(E)}}\right)p_j$.
Now, the new optimization program can be written as:
\begin{align}
&\max_{(\{\eta_i\}_{i\in \mathcal{C}},\{p_j\}_{j=1}^N)} \sum_{i\in \mathcal{C}} \eta_i \sum_{j}\left(\frac{\Omega_j^{(E)}-\Omega_j^{(i)}}{\Omega_j^{(i)}\Omega_j^{(E)}}\right)p_j   \label{eq:opt2}\\
\text{s.t.} \ \  &\sum_{i\in \mathcal{C}} \eta_i \sum_{j=1}^N p_j\leq P_T \nonumber\\
		&\sum_{i\in \mathcal{C}} \eta_i= 1   \nonumber 	
\end{align}
or in more compact form:
\begin{align}
&\max_{(\mathbf{\eta},\mathbf{p})} \ \ \mathbf{\eta}^T \mathbf{C}\mathbf{p}   \label{eq:opt3}\\
&\text{s.t.} \ \  \mathbf{\eta}^T \mathbf{I}\mathbf{p}\leq {P_T} \nonumber\\
		&(\mathbf{1})^T \mathbf{\eta}= 1   \nonumber \\
		& \mathbf{p}\geq \mathbf{0}   \nonumber 	
\end{align}
where $\mathbf{\eta}=[\eta_1 \ .\  .\ .\ .\ \eta_M]^T$, $\mathbf{C}=[\mathbf{c^{(1)^T}} \ .\ .\ .\ .\ \mathbf{c^{(M)^T}}]$, $\mathbf{c^{(i)}}=[\frac{\Omega_1^{(E)}-\Omega_1^{(i)}}{\Omega_1^{(i)}\Omega_1^{(E)}}\ .\ .\ .\ .\ \frac{\Omega_N^{(E)}-\Omega_N^{(i)}}{\Omega_N^{(i)}\Omega_N^{(E)}}]^T$, $\mathbf{I}$ is the $N\times N$ identity matrix and $\mathbf{p}=[p_1\ .\ .\ .\ .\ p_N]^T$ \\
The program (\ref{eq:opt3}) is a mixed binary integer linear programming problem. We solve it through decomposition or separation approach. For a selected node, program (\ref{eq:opt3}) becomes
\begin{align}
&\max_{\mathbf{p}} \ \ (\mathbf{c})^T\mathbf{p}   \label{eq:opt4}\\
&\text{s.t.} \ \  (\mathbf{1})^T\mathbf{p}\leq P_T \nonumber\\
		& \mathbf{p}\geq \mathbf{0}   \nonumber 	
\end{align}
The problem (\ref{eq:opt4}) is a linear program. The feasible set/region (due to constraints 1 and 2 of (\ref{eq:opt4})) constitutes a part of $l_1$-unit ball extended to $P_T$. Thus, $p_j$, $\forall j$ takes a value which is either zero, or, positive, as shown in Fig. $2$. 

\begin{figure}
\label{Feasibleset}
\begin{center}
\includegraphics[width=6cm]{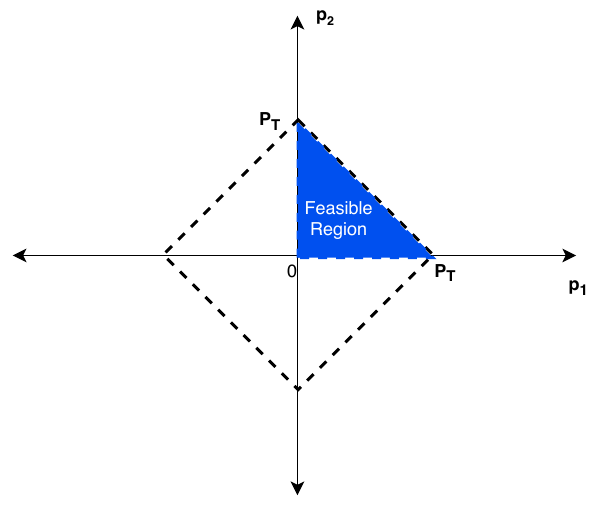}
\caption{The $l_1$-unit ball extended to $P_T$ for $N=2$, shaded area shows the feasible region.}
\end{center}
\end{figure}

 Let $\nu = \{\mathbf{v_j}\}_{j=1}^N$ be the set of vertexes of the feasible set. Then, a common approach for optimal solution to problem (\ref{eq:opt4}) is: 
\begin{align}
\label{Eq:sol of LP}
\mathbf{p^*}= \mathbf{v_j^*}= arg\max_{j} \ \  (\mathbf{c})^T\mathbf{v_j} 
\end{align}
Now, by closely inspecting the solution in Eq. (\ref{Eq:sol of LP}), we can see that this solution makes the OFDM system (a multi-carrier system) a single-carrier system by allocating zero power to all but one sub-carriers. Simply speaking, this solution is not desired due to two main reasons. First, it changes a multi-carrier system into a single-carrier system. Second, it makes the argument of the $\log$ function large, and we know that $\log$ compress large values. Therefore, we provide an alternate optimal solution to problem (\ref{eq:opt4}) by carefully studying the nature of the problem.\\ 

The objective function in program (\ref{eq:opt4}) is an inner product which we can re-write as $\langle \mathbf{c}, \mathbf{p}\rangle$. We know that co-linear vectors maximize the inner product but the problem is to find such a vector $\mathbf{p}$ which is co-linear to $\mathbf{c}$ and it meets the constraints $1$ and $2$ of program (\ref{eq:opt4}) as well. Note that $\mathbf{c}$ can be either zero, or, all positive, or, all negative. When $\mathbf{c}$ is negative or zero, then the optimal solution is to allocate zero power to all the sub-carriers. When $\mathbf{c}$ is positive, then the solution is to choose such a vector which is co-linear to $\mathbf{c}$ and meets the constraint 1. For this, first we normalize the $\mathbf{c}$ to have a unit $l_1$-norm, i.e., $\mathbf{\hat{c}}=\frac{\mathbf{c}}{\Vert \mathbf{c}\Vert_1}$. Then, the optimal solution to the problem (\ref{eq:opt4}) is:
\begin{align}
\mathbf{p^*}=P_T\otimes\mathbf{\hat{c}}, \label{eq:optLP}
\end{align}  
 where $\otimes $ denotes the Kronecker product. Now, putting $\mathbf{p}^*$ back to program (\ref{eq:opt3}), we are left with the following problem:
\begin{align}
&\max_{\bf{\eta}}  \ \ \eta^T \bf{C}\bf{p}^*   \label{eq:opt5}\\
\text{s.t.} \ \  & (\mathbf{1})^T \mathbf{\eta}= 1   \nonumber	 	
\end{align}
The optimal solution to (\ref{eq:opt5}) is:
\begin{align}
\label{eq:opt6}
\eta^* &=\begin{cases}
\eta_i=1, &when \ \ i=index\left( \Vert\mathbf{C}\mathbf{p^*}\Vert_{\infty}\right) \\
\eta_i=0 &\ \mathrm{else}
\end{cases}
\end{align}
This completes the description of the proposed sub-optimal scheme. 
The proposed method (when run on the first hop) is summarized in Algorithm 2.

\section{Optimal, sub-optimal and ML solutions to the NSPA problem in multipath UWA channel}
In this section, we discuss the case where we have a multipath UWA channel on each hop (between the sensor node $A_0$ and the sink node). With this, we derive the optimal power allocation for both (optimal and sub-optimal) schemes.

\subsection{Optimal scheme}
The secrecy rate of $i^{th}$ node is as follows: 
 \begin{equation}
 \label{eq:Cd}
SR_{MP}^i =  \bigg( \Delta f (\sum_{j=1}^{N} \log_2(1+\gamma_j^i(k))-\log_2(1+\gamma_j^E(k))) \bigg)^{+},
\end{equation}
where $\gamma_j^i(k)=\frac{p_j\vert H_j^i\vert^2}{\int_{(j-1)\Delta f}^{j\Delta f}N(f) df}$, $\gamma_j^E(k)=\frac{p_j\vert H_j^E\vert}{\int_{(j-1)\Delta f}^{j\Delta f}N(f) df}$. 

The formulated optimization program for the multipath UWA channel is given as: 
\begin{align}
&\max_{(\{\eta_i\}_{i\in \mathcal{C}}, \{p_j\}_{j=1}^N)} \sum_{i\in \mathcal{C}} \eta_i \mathrm{SR_{MP}}^{(i)}(\{p_j\}_{j=1}^{N})   \label{eq:opt_mp}\\
\text{s.t.} \ \  &\sum_{i\in \mathcal{C}}  \eta_i \sum_{j=1}^N p_j\leq P_T \nonumber\\
		&\sum_{i\in \mathcal{C}} \eta_i= 1   \nonumber  
\end{align}
The program (\ref{eq:opt_mp}) is similar to (\ref{eq:opt}) (i.e in constraints and nature (mixed binary integer program)) but the difference lies in the objective function. We adopt the similar mechanism as used in Section IV, to solve (\ref{eq:opt_mp}). For the sake of brevity, we omit the steps, while the optimal power that we get is given as:
 \begin{align}
\label{eq:optpower_mp}
p_j^*=\left(\frac{-b_j+\sqrt{b_j^2-4a_jc_j}}{2a_j}\right)^+,
\end{align}
where $a_j= \bar{H_j^i}\bar{H_j^E} $,\ \ $b_j=\bar{H_j^i}+\bar{H_j^E} $,  $c_j=1-\frac{\bar{H_j^i}-\bar{H_j^E}}{\ln(2) \lambda}$, $\bar{H_j^i}=\frac{\vert H_j^i\vert^2}{\int_{B_j(f)}N(f) df}$ and $\bar{H_j^E}=\frac{\vert H_j^E\vert^2}{\int_{B_j(f)}N(f) df}$.
The steps for node selection and solving dual problem are similar to section IV.
\subsection{Sub-optimal scheme}
The objective function for sub-optimal scheme is $\sum_{j=1}^N (\bar{H_j^i}-\bar{H_j^E})p_j  $. The formulated optimization program with the mentioned objective function is given as: 
\begin{align}
&\max_{(\mathbf{\eta},\mathbf{p})} \ \ \mathbf{\eta}^T \mathbf{C_{MP}}(\mathbf{p})   \label{eq:opt_sub_mp}\\
&\text{s.t.} \ \  \mathbf{\eta}^T \mathbf{I}(\mathbf{p})\leq {P_T} \nonumber\\
		&(\mathbf{1})^T \mathbf{\eta}= 1   \nonumber \\
		& \mathbf{p}\geq \mathbf{0}   \nonumber 	
\end{align}
The program (\ref{eq:opt_sub_mp}) is similar to (\ref{eq:opt3}) but the difference lies in $\mathbf{C_{MP}}$. Here, $\mathbf{C_{MP}}=[\mathbf{c_{MP}^{(1)^T}} \ .\ .\ .\ .\ \mathbf{c_{MP}^{(M)^T}}]$, $\mathbf{c_{MP}^{(i)}}=[(\bar{H_1^i}-\bar{H_1^E}) \ .\ .\ .\ .\ (\bar{H_N^i}-\bar{H_N^E})]^T$. To solve (\ref{eq:opt_sub_mp}), we repeat the steps of section V to get the optimal solution for power allocation: 
\begin{align}
\mathbf{p^*}=P_T\otimes\mathbf{\hat{c}_{MP}}, \label{eq:optLP_mp}
\end{align}
where $\mathbf{\hat{c}_{MP}}=\frac{\mathbf{c_{MP}}}{\Vert \mathbf{c_{MP}}\Vert_1}$, while the procedure for node selection is same as in Section V.

\subsection{ML-based optimization}
\begin{figure*}
\begin{center}
\includegraphics[width=22cm]{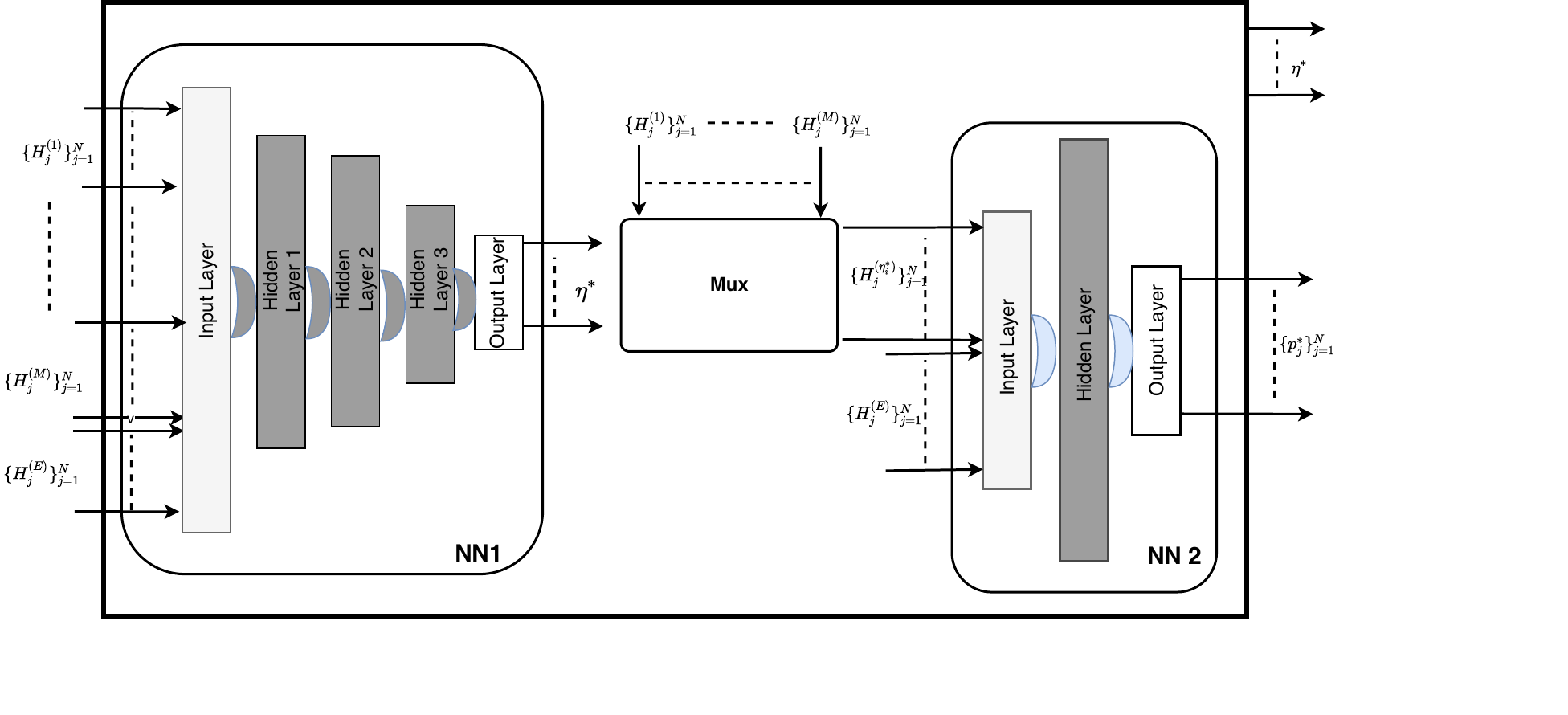}
\caption{The proposed Learning-based architecture to solve the NSPA problem.}
\label{fig:LBPM}
\end{center}
\end{figure*}

Though the proposed sub-optimal scheme is a one-shot method, but simulation results reveal that its performance is far below than the optimal scheme in multipath UWA channel scenario.  Additionally, the proposed sub-optimal scheme is out-performed by the constant-power allocation scheme, for high power budgets. Therefore, in this sub-section, we also solve the NSPA problem at hand via machine learning techniques (specifically, neural networks), which allows us to reduce the time-complexity of the NSPA problem. By closely inspecting our formulated optimization program, we notice that it is a combination of classification problem and regression problem. That is, the selection of the helper/relay node is a classification problem (with $M$ classes), while the power loading over sub-carriers is a regression problem. So, we use two NNs to solve the NSPA problem (\ref{eq:opt_mp}). The proposed learning-based methodology with two NNs is shown in Fig. \ref{fig:LBPM}.   
\subsubsection{NN1} The NN1 structure is shown in Fig. \ref{fig:LBPM}. We use NN1 for the purpose of classification (node selection). Here, we set up an NN with $5$ layers (an input layer, three hidden layers, and an output layer). At the input layer, we have total numbers of neurons equal to the total sub-carriers times the total number of nodes (i.e., $(M+1)N$). Then we successively reduce the total number of neurons by factor of $2$ at each layer till we reach the output layer. Thus, we have a total of $M$ neurons at the output layer. We use Rectified Linear Unit (ReLU) as  an activation function at all the hidden layers and \textit{softmax} at the output layer for obtaining output probability distribution from un-normalized output. ReLU can be expressed as: 
\begin{align}
\text{ReLu}(z) &=\begin{cases}
z, &when \ \ z>0 \\
0 &\ \mathrm{else}
\end{cases}
\end{align}
where $z$ is considered as input to neurons at hidden layers while the \textit{softmax} is given as
\begin{align}
\text{s}_i=\frac{e^{y_i}}{\sum_{m=1}^M (e^{y_m})},
\end{align}
where $y_i$ is the un-normalized output. The Cross Entropy (CE) loss function at the output layer is given as
\begin{align}
\text{CE}=-\sum_{m=1}^Mt_mlog(s_m),
\end{align}
where $t_m$ is the true label of class $m$ where $m\in \{M\}$.

\subsubsection{NN2} The architecture of NN2 is shown in Fig. \ref{fig:LBPM}. Here, we choose an NN with three layers (i.e., an input layer, a hidden layer, and an output layer). Remember that we use NN2 to solve the regression problem (i.e., power loading over the OFDM sub-carriers). We have $2N$ neurons at the input layer, $16N$ neurons at the hidden layer, and $N$ neurons at the output layer. We chose ReLU as an activation function at the hidden layer to enforce the individual power constraint (i.e. $p_j\geq 0$). Now, to ensure the sum power constraint, we play with the Mean Squared Error (MSE) loss function of NN2 which can be expressed as:
\begin{align}
\text{L}=&\frac{\sum_{n=1}^{B_T}(Y_{act}(n)-Y_{pred}(n))^2}{({B_T})}+ \\
&\frac{\sum_{n=1}^{B_T}(max(0,{P_T-\sum_{j}Y_{pred}(n)^{j}}))}{({B_T})}, \nonumber
\end{align}
where $Y_{act}$ is the true label obtained through Eq. (\ref{eq:optpower_mp}), while $Y_{pred}$ is obtained through NN2. $B_T$ is the total number of samples in a batch over which the loss is computed. This loss function is designed by keeping two things in mind: 1) it should decrease the error between actual and predicted value (first term of loss function) 2) it should also keep predicted power budget within limits of specified $P_{T}$. This constraint is imposed by the second term of the loss function.
\subsection{Computational Complexity}
Let $TI$ be the number of  total iterations that optimal scheme takes to converge. To compute $p_j$ we need $7$ multiplications  $\forall j$ then the total computational complexity of joint optimal scheme is $O(7(M)(TI)(N))$, while the computational complexity for the sub-optimal scheme is $O(2(M)(N))$.\\
The computational complexity of an NN is the total number of parameters that it needs to learn. The number of parameters in turn depend on the number of neurons used at the input, hidden and the output layers. Let $N_1^I$, $N_{1}^{H1}$, $N_{1}^{H2}$, $N_1^{H3}$ and $N_1^O$ be the number of neurons at the input, hidden and the output layers respectively for NN1, and $N_2^I$, $N_2^{H}$, $N_2^{O}$ be the number of neurons at the input, hidden and the output layers respectively for NN2. Then, the computational complexity of the proposed learning-based scheme is: $O\left(N_1^{H1}(N_1^I+N_1^{H2})+N_1^{H3}(N_1^{H2}+N_1^{O})+N_2^H(N_2^I+N_2^O)\right)$. Adopting to our case, the computational complexity becomes: $O((MN)(\frac{21MN+4M}{32})+\frac{36N}{M})$.

\section{Simulation Results}
\subsection{Simulation Setup} 

The simulations were performed in MATLAB and Python. We deploy $M$ number of legitimate nodes, and an Eve node according to a uniform distribution in a (vertical) square region of area $500*500$ m$^2$,
under the water. We place the sink node on the top of water surface. 
We assume an OFDM system whose parameters are mentioned in TABLE I (the choice of values for these parameters was guided by \cite{Stojanovic:1ACMIWUN:2006}).
 \begin{table}[h]                           
 \centering
    \begin{tabular}{|c|c|c|}
    \hline
\textbf{Parameter name}    &  \textbf{Notation}  & \textbf{Value} \\ \hline

     Total Number of sub-carriers    &   $N$ & $32$               \\ \hline
     Total Bandwidth &   $B$ &    $6$ kHz         \\ \hline 
      Sub-carrier Bandwidth &   $\Delta f$ &    $\frac{B}{N}$ kHz         \\ \hline
     Frequency range    & $f_N$-$f_1$  & $15-9$ kHz \\ \hline
    Total multi paths          & L   & $3$          \\ \hline
    Spreading factor   &  $\nu$ &       $1.5$      \\ \hline
   Speed of sound   &   $v$ & $1500$ m/s             \\ \hline
    Experimental constants     &  $N_1$ and $\tau$  & $50$ dB and $18$ dB    \\ \hline

    \end{tabular}
    \caption{Simulation Parameters}
    \label{tab:msg1}                            

\end{table}
We generate channel gains by considering three paths (i.e. direct path, surface reflection and bottom reflection) in Eq. (\ref{CGinMP}). We assume that each sensor node knows the distance (channel gains) of Eve in LoS (multipath) scenario from itself.\footnote{This assumption is inline with the previous works which perform secrecy rate analysis of the underwater/terrestrial communication systems \cite{Wang:ICST:2017},\cite{Huang:SJ:2016}.} The secrecy rate SR plotted in each of the following figures is the minimum secrecy rate among all the hops (after reaching data to sink node). In other words, $\text{SR}=\min \{\text{SR}_1,...,\text{SR}_k,...,\text{SR}_K\}$ where SR$_k$ is the secrecy rate obtained by solving the optimization program (\ref{eq:opt}) ((\ref{eq:opt_mp})) for LoS (multipath) at $k$-th hop (assuming that there are $K$ hops in total. Similar goes for other schemes as well. Furthermore, to investigate the importance of node selection in the results below, the benchmark used to assess the performance of proposed schemes is the classical Depth-Based Selection (DBS) scheme for data forwarding \cite{Yan:ICRN:2008}. 
Briefly speaking, for a sender node with data, the DBS scheme selects at each hop a relay node for data forwarding that has maximum depth among all the candidate helper nodes with depth greater than the depth of the sender node.\\

{\it Generation of data set for training of NN1}: We generate $1e^5$ realizations of uniform distribution for $M+1$(including Eve) nodes in the above-mentioned region. Every time we select the first node as Tx and then we execute the solution of problem (\ref{eq:opt_mp}) to compute the optimal forwarding node or labels for NN1. Finally, we have $(M+1)N \times 1e^5$ dimensional matrix as input and $M \times 1e^5$ as output for training NN1. Entire data is not sent at once to the neural network, we use mini batch training approach. \\
{\it Generation of data set for training of NN2}: We save the channel gains of every node from $A_0$ and the corresponding labels (i.e. $p_j^\ast$ obtained through Eq. (\ref{eq:optpower_mp})) as well. But we considered  first $N \times 1e^5$ channel gains  with the channel gains of Eve sufficient to find relation between channel  gains and  $p_j^\ast \ \ \forall j$. So, we have  $2N \times 1e^5$ dimensional matrix as input and $N\times 1e^5$ dimensional matrix as output to train NN2.\\
For both networks NN1 and NN2, learning rate of $0.01$ is used as an initial value and it is decreased by the factor of $10$ when training loss plateaus. Adam optimizer is used to train both networks. Total data is divided into training and validation sets by the fraction of $0.8$ and $0.2$ respectively. For both networks we keep batch size $B_T=50$ and Epochs$=500$.


\vspace{2mm}
\subsection{Simulation Results}

\begin{figure}
\begin{center}
\includegraphics[width=9.5cm]{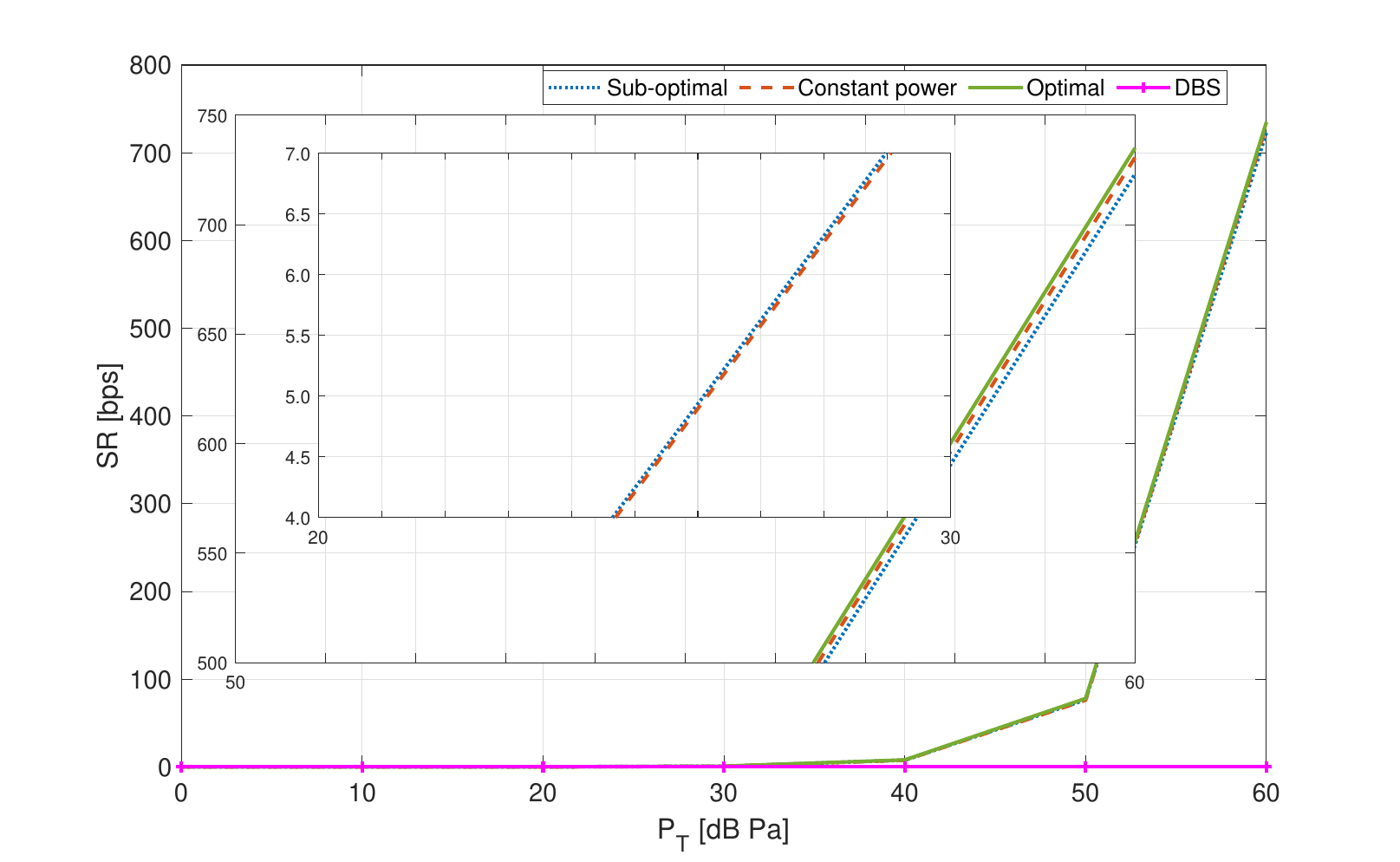}
\caption{The impact of transmit power budget of the sender nodes on the secrecy rate for the scenario of LoS UWA channel.}
\label{fig:sc-vs-pt}
\end{center}
\end{figure}

Fig. \ref{fig:sc-vs-pt} studies the impact of transmit power budget on secrecy rate achieved for the proposed optimal, sub-optimal, constant power and the DBS schemes, under the scenario of LoS UWA channel. For Fig. \ref{fig:sc-vs-pt}, we set $M=10$ and plot the average of $1000$ random realization of node deployment. We make the following observations: i) The secrecy rate is an increasing function of the total power budget $P_T$. ii) The proposed optimal scheme out-performs all the other schemes with a slight margin, while the sub-optimal scheme outperforms the constant power scheme with slight margin below $P_T=50$ dBPa, and vice versa. iii) Constant/equal power allocation over the sub-carriers is near-to-optimal solution for the case of LoS UWA channel. iv) The DBS scheme lies on the x-axis throughout, which shows the importance of node selection for enhancing the secrecy rate. We obtained a secrecy rate that is identically zero for optimal, sub-optimal and constant power schemes in case of DBS scheme, that is why, one can see only one curve for the DBS scheme. Furthermore, by closely inspecting the performance of DBS, we came to know that there is at least one hop which results in negative or zero secrecy rate. 
 


\begin{figure}
\begin{center}
\includegraphics[width=9.5cm, height=6cm]{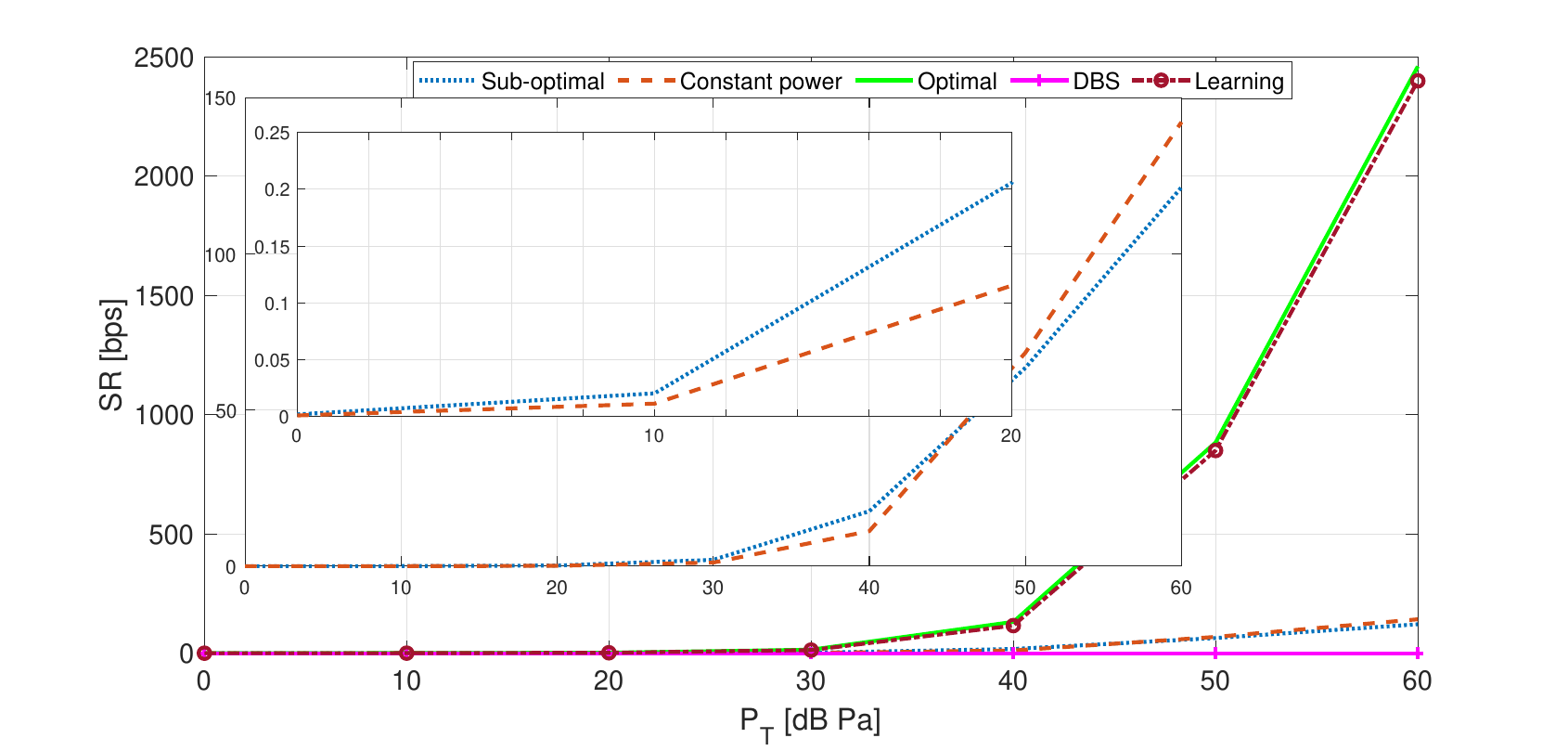}
\caption{The impact of transmit power budget of the sender nodes on the secrecy rate for the scenario of a multipath UWA channel.}
\label{fig:sc-vs-pt_MP}
\end{center}
\end{figure}

Fig. \ref{fig:sc-vs-pt_MP} studies the impact of transmit power budget on secrecy rate achieved for the proposed optimal, sub-optimal, constant Power, learning-based and the DBS schemes, under the scenario of multipath UWA channel. Here, one can notice the similar trends as in Fig. \ref{fig:sc-vs-pt}. Nevertheless, there are some important differences that are as follows. The importance of the power allocation among the OFDM sub-carriers is quite prominent here as one can see the significant gap between the curve for the optimal scheme and the rest of the schemes; moreover, the gap increases as $P_T$ increases. This gap motivated us to find another sub-optimal scheme that should perform very close to optimal scheme, which is precisely the proposed learning-based scheme. 

\begin{figure}
\begin{center}
\includegraphics[width=9.5cm,height=6cm]{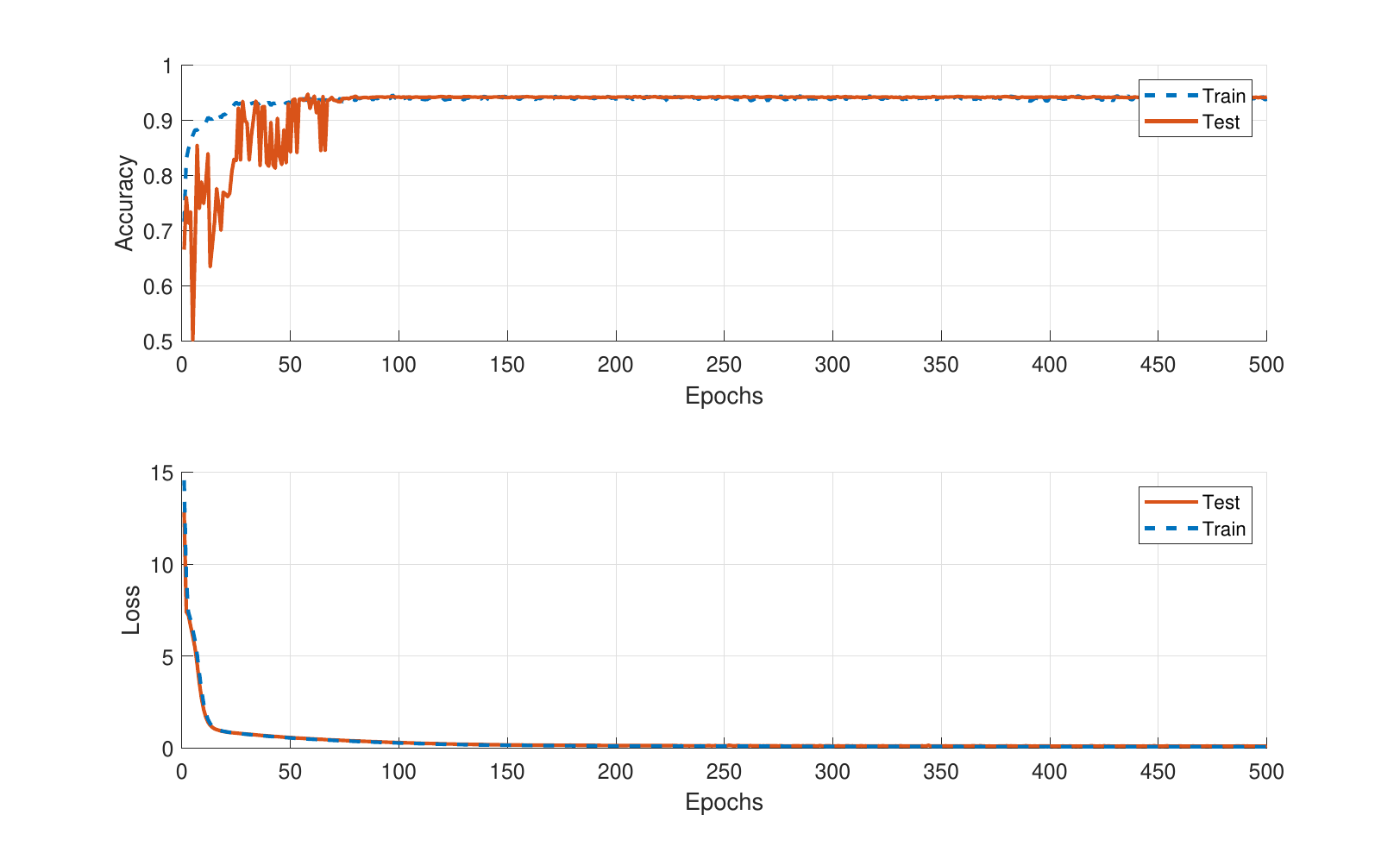}
\caption{Performance of NN1 (top sub-plot) and NN2 (bottom sub-plot). One may observe that $150$ Epochs are good enough for training.}
\label{fig:nn1}
\end{center}
\end{figure}
Fig. \ref{fig:nn1} shows the strength of the proposed learning-based scheme. One can clearly observe that the trained models are not under-fit or over-fit. Specifically, the top sub-plot shows the accuracy of learning based classifier (i.e. NN1) which approaches $95\%$  for both test and training data. On the other hand, the bottom sub-plot shows the MSE loss for test and training data of NN2, which nearly touches zero.

\begin{figure}
\begin{center}
\includegraphics[width=9.5cm,height=6cm]{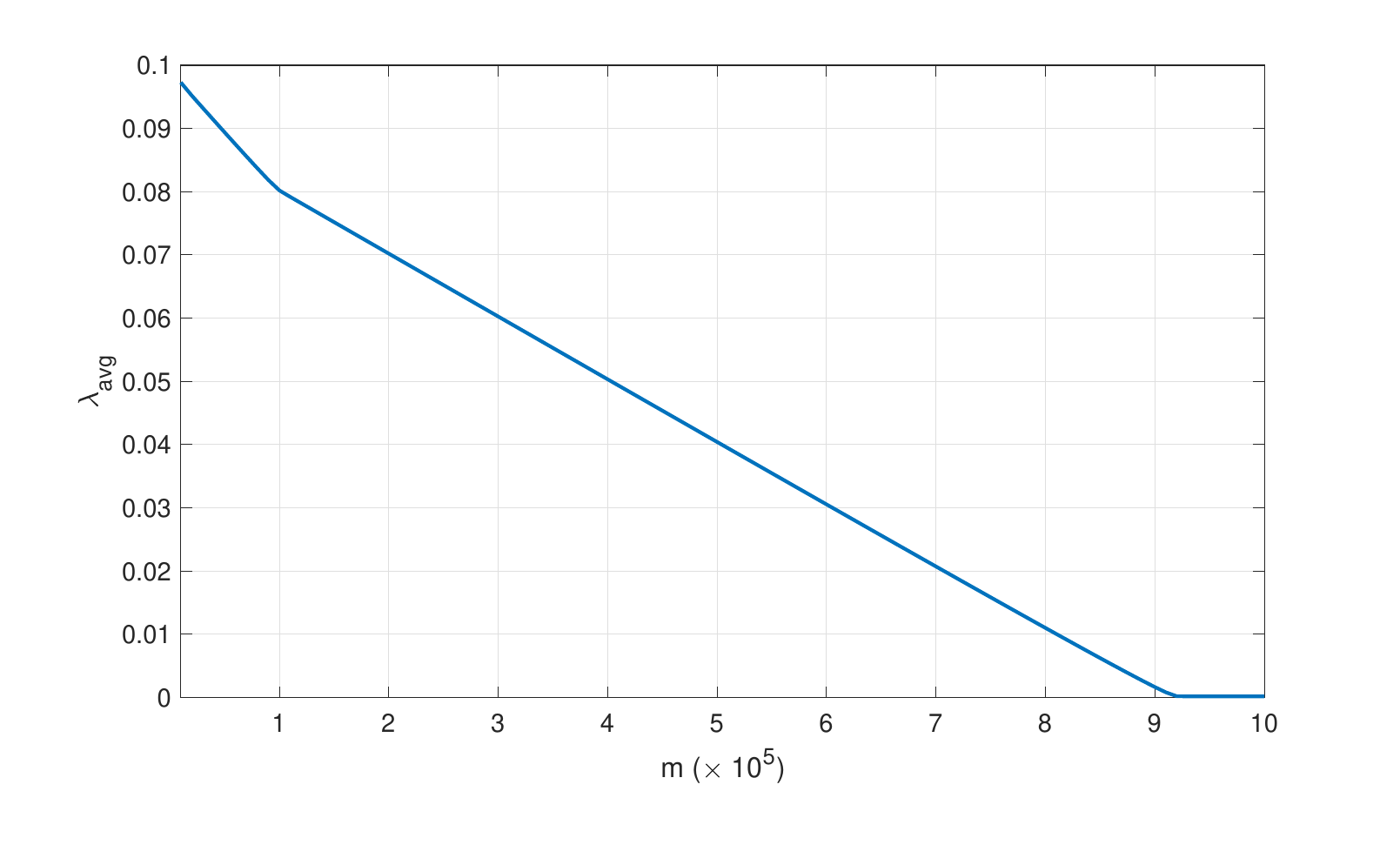}
\caption{$\lambda_{avg}$ vs. number of iterations}
\label{fig:lambda}
\end{center}
\end{figure}
Finally, Fig. \ref{fig:lambda}. which aims to find out the value of $TI$. Fig. \ref{fig:lambda} is generated after $100$ uniform random realizations of the dual variable $\lambda$ between $0$ and $1$, and then taking its average. One can see that $\lambda_{avg}$ gets constant at roughly around $0.9$ million iterations.
\begin{tiny}
\begin{table}
\centering
 \begin{tabular}{||c c c c ||} 
 \hline
 $N$ & Optimal & Sub-optimal & Learning-based \\ [0.5ex] 
 \hline\hline
 $32$ & $2.01*10^{10}$ & $6400$ &$6.8*10^{6}$ \\ 
 \hline
$64$ & $4.03*10^{10}$ & $12800$ & $2.7*10^{7}$ \\
 \hline
 $128$ & $8.06*10^{10} $ & $25600$ & $1.1*10^{8}$ \\
 \hline
 $256$ & $1.61*10^{11}$ & $51200$ & $4.3*10^{8}$ \\
 \hline
 $512$ & $3.22*10^{11} $& $102400$ & $1.7*10^{9}$ \\
 \hline
 $1024$ & $6.45*10^{11} $& $204800$& $6.9*10^{9}$\\ [1ex]
\hline
\end{tabular}
\caption{Tabular comparison of computational complexity of Optimal, Sub-Optimal and Learning-based schemes with increase in number of OFDM sub-carriers}
\end{table}
\end{tiny}
\begin{tiny}
\begin{table}
\centering
 \begin{tabular}{||c c c c||} 
 \hline
 $M$ & Optimal & Sub-optimal & Learning-based \\ [0.5ex] 
 \hline\hline
 $20$ & $1.29*10^{11}$ & $40960$ &$313051136$ \\ 
 \hline
$40$ & $2.58*10^{11}$ & $81920$ & $1.13*10^{9}$ \\
 \hline
 $60$ & $3.87*10^{11} $ & $122880,$ & $2.51*10^{9}$ \\
 \hline
 $80$ & $5.16*10^{11}$ & $163840$ & $4.4*10^{9}$ \\
 \hline
 $100$ & $6.45*10^{11} $& $204800$ & $6.92*10^{9}$ \\
 [1ex]
\hline
\end{tabular}
\caption{Tabular comparison of computational complexity of Optimal, Sub-Optimal and Learning-based schemes with increase in number of nodes}
\end{table}
\end{tiny}
TABLE II \& III show the computational complexity of the three proposed schemes: optimal scheme, sub-optimal scheme, and the learning-based scheme. We keep $M = 100$ for TABLE II results and $N = 1024$ for TABLE III results. One can see that even for the extreme cases, the computational complexity of the learning-based scheme is negligible compared to the optimal scheme. The computational complexity of the sub-optimal scheme is indeed very low compared to the other two schemes but it also results in low secrecy rate as seen in Fig. \ref{fig:sc-vs-pt_MP}.

\section{Conclusion} \label{sec:con}

This work presented the first study focused on an eavesdropping attack on an OFDM-based, multi-hop UWASN from the perspective of resource allocation (helper/relay node selection and power allocation among the OFDM sub-carriers). Specifically, we performed the joint optimal and sub-optimal node selection for data forwarding and power allocation for maximizing the secrecy rate at each hop, for the scenario of an LoS UWA channel. The analysis was also extended to the scenario of a multipath UWA channel where we also solved the joint optimization program at hand via a novel learning-based method. Simulation results revealed that given a total power budget of $60$ dBPa, the optimal scheme yielded a secrecy capacity of roughly $750$ ($2400$) bps in LoS (multipath) UWA channel. This prompts us to deduce that the multipath UWA channel--being more random in nature—favors the cause of secrecy capacity more compared to the LoS UWA channel. The proposed learning-based scheme, on the other hand, performed near optimal coupled with the additional benefit of having reduced computational complexity.\\

Some comments about the proposed (optimal and sub-optimal) schemes are in order. The proposed schemes are scalable with increase in network dimension as they simply need to be re-run on each additional hop; therefore, increasing $M$ does not change the analysis. Furthermore, in case of multiple malicious nodes, one heuristic/sub-optimal approach will be to find out the eavesdropping link with maximum leakage, and then proceed with the proposed schemes. Another viable approach in such situation will be to use alternating optimization or max-min optimization for enhancing the overall secrecy rate. We leave this problem of tackling eavesdropping attack by multiple malicious nodes for future work.\\

Some other potential ideas for future work are as follows. One could study the impact of jamming attack on a UWASN from one or more malicious nodes and suggest viable countermeasures. One could implement the concept of artificial noise generation in a UWASN at the transmitter, receiver or a friendly relay in order to equivocate a passive eavesdropper node to enhance the secrecy rate of the UWASN under consideration.

\footnotesize{
\bibliographystyle{IEEEtran}
\bibliography{main}
}

\vfill\break

\end{document}